\documentclass[journal=nalefd, manuscript=article]{achemso}

\usepackage{epsfig}
\usepackage{graphicx}
\usepackage{amsmath}

\usepackage[usenames]{color}

\author{Hanghui Chen}
\email{hc2650@columbia.edu}
\affiliation
{Department of Physics, Columbia University, New York, NY 10027, USA}
\alsoaffiliation
{Department of Applied Physics and Applied Mathematics, Columbia University, New York, NY 10027, USA}
\alsoaffiliation
{Department of Physics, Yale University, New Haven, CT 06511, USA}
\alsoaffiliation
{Department of Applied Physics, Yale University, New Haven, CT 06511, USA}
\alsoaffiliation
{Center for Research on Interface Structures and Phenomena (CRISP), Yale University, New Haven, CT 06511, USA}
\author{Qiao Qiao}
\affiliation
{Department of Physics, University of Illinois at Chicago, Chicago, IL 60607, USA}
\author{Matthew S. J. Marshall}
\affiliation
{Department of Applied Physics, Yale University, New Haven, CT 06511, USA}
\alsoaffiliation
{Center for Research on Interface Structures and Phenomena (CRISP), Yale University, New Haven, CT 06511, USA}
\author{Alexandru B. Georgescu}
\affiliation
{Department of Physics, Yale University, New Haven, CT 06511, USA}
\alsoaffiliation
{Center for Research on Interface Structures and Phenomena (CRISP), Yale University, New Haven, CT 06511, USA}
\author{Ahmet Gulec}
\affiliation
{Department of Physics, University of Illinois at Chicago, Chicago, IL 60607, USA}
\author{Patrick J. Phillips}
\affiliation
{Department of Physics, University of Illinois at Chicago, Chicago, IL 60607, USA}
\author{Robert F. Klie}
\affiliation
{Department of Physics, University of Illinois at Chicago, Chicago, IL 60607, USA}
\author{Frederick J. Walker}
\affiliation
{Department of Applied Physics, Yale University, New Haven, CT 06511, USA}
\alsoaffiliation
{Center for Research on Interface Structures and Phenomena (CRISP), Yale University, New Haven, CT 06511, USA}
\author{Charles H. Ahn}
\affiliation
{Department of Applied Physics, Yale University, New Haven, CT 06511, USA}
\alsoaffiliation
{Center for Research on Interface Structures and Phenomena (CRISP), Yale University, New Haven, CT 06511, USA}
\author{Sohrab Ismail-Beigi}
\affiliation
{Department of Applied Physics, Yale University, New Haven, CT 06511, USA}
\alsoaffiliation
{Center for Research on Interface Structures and Phenomena (CRISP), Yale University, New Haven, CT 06511, USA}

\title{Reversible modulation of orbital occupations via an
interface-induced polar state in metallic manganites}


\begin{document}

\begin{abstract}
The breaking of orbital degeneracy on a transition metal cation and
the resulting unequal electronic occupations of these orbitals provide
a powerful lever over electron density and spin ordering in metal
oxides. Here, we use {\it ab initio} calculations to show that
\textit{reversibly} modulating the orbital populations on Mn atoms
can be achieved at ferroelectric/manganite interfaces by the
presence of ferroelectric polarization on the nanoscale. The
change in orbital occupation can be as large as 10\%, greatly
exceeding that of bulk manganites. This reversible orbital splitting
is in large part controlled by the propagation of ferroelectric polar
displacements into the interfacial region, a structural motif absent
in the bulk and unique to the interface. We use epitaxial thin film
growth and scanning transmission electron microscopy to verify this
key interfacial polar distortion and discuss the potential of
reversible control of orbital polarization via nanoscale ferroelectrics.
\end{abstract}

A key characteristic of transition metal oxides is the presence of
electronically active $d$ orbitals on the transition metal
cations~\cite{Dagotto, Benckiser-NatMater-2011}. This degree of
freedom creates a rich variety of behaviors, and a large area in
materials science and technology focuses on understanding and
controlling these properties (e.g. magnetism, superconductivity,
ferroelectricity, etc.)~\cite{Tokura-NatMat-2008}. Perovskite complex
oxides form a large subset of such oxides. For a transition metal in a
cubic perovskite, crystal fields split its five $d$ orbitals into a
lower energy three-fold degenerate $t_{2g}$ manifold ($d_{xy}$,
$d_{xz}$, $d_{yz}$) and a higher energy two-fold degenerate $e_g$
manifold ($d_{3z^2-r^2}$, $d_{x^2-y^2}$)
~\cite{Tokura-Science-2000}. These degeneracies can be further
removed, for example, by Jahn-Teller distortions, in order to create
unequal electronic occupancies within each originally degenerate
manifold~\cite{Goodenough-ARMC-1998}.  The resulting charge anisotropy
can in turn affect electronic transport and magnetic
ordering~\cite{Fang-PRL-2000}. Hence, controlling the energies of the
$d$ orbitals tailors the physical properties of metal oxides in the
bulk as well as at surfaces and
interfaces~\cite{Chakhalian-Science-2007,
  Rata-PRL-2008,Tebano-PRL-2008,Huijben-PRB-2008, Salluzzo-PRL-2009,
  Yu-PRL-2010}. A classic example is provided by manganites, where the
$e_g$ orbitals are active in transport and
magnetism~\cite{Salamon-RMP-2001}. The energetic ordering of the
$e_g$ orbitals on each Mn site as well as neighboring Mn sites
profoundly affects the ground state magnetic
properties~\cite{Solovyev-PRL-1996, Molegraaf-AdM-2009, Lu-APL-2012,
  Garcia-Science-2010, Burton-PRB-2009, Burton-PRL-2011,
  Vaz-PRL-2010}.

Although structural distortions (e.g., Jahn-Teller or GdFeO$_3$
distortions) are common for bulk perovskite manganites, they only
weakly remove orbital degeneracy~\cite{Salamon-RMP-2001}. With the
development of epitaxial thin film growth techniques, it is possible
to remove orbital degeneracy through strain-induced Jahn-Teller-like
distortions. Tensile (compressive) strain modifies the crystal field
so as to favor the in-plane orbital $d_{x^2-y^2}$ (out-of-plane
orbital $d_{3z^2-r^2}$)~\cite{Tebano-PRL-2008, Huijben-PRB-2008}.
However, utilizing strain is a static approach to tailoring the
desired orbital configuration~\cite{Sadoc-PRL-2010}.

In this Letter, we describe an approach that utilizes
nanoscale ferroelectrics and enables \textit{reversibly}
modulating orbital occupations at La$_{1-x}$Sr$_x$MnO$_3$ (LSMO,
$x$=0.2 for the current study) interfaces. We begin with first
principles calculations which show that switching the ferroelectric
polarization at a (001) ferroelectric/manganite interface can modulate
the atomic-scale structure, change the electronic distribution at the
interface, and split the orbital degeneracy of the interfacial Mn
$e_g$ levels. Furthermore, the sign of the splitting is opposite for
the two different ferroelectric polarizations and the resulting
changes in orbital occupancies can be as large as 10\%. Then we use
experimental growth and characterization to demonstrate the predicted
key interfacial structural distortion that underlies the orbital
splitting.

Our theoretical calculations are based on density functional theory
with a plane wave basis set and ultrasoft pseudopotentials as
implemented in the quantum-espresso package~\cite{QE}. The technical
details can be found in the Supplemental Material~\cite{supplement}.
We choose BaTiO$_3$ as the ferroelectric prototype in order to have a
direct comparison to our experiments. However, our qualitative
predictions are independent of the choice of ferroelectrics, as
explained below.  We use an in-plane $c(2\times2)$ cell to incorporate
rotations and tiltings of MnO$_6$ octahedra in the manganites. A
periodic electrostatic boundary condition is imposed on the supercell
which includes the material slabs and the vacuum.  The interface
considered here is a BaO/MnO$_2$ ferroelectric/manganite (001)
interface. We realize this interface experimentally by using molecular
beam epitaxy to deposit an integer number of unit cells of LSMO on a
Nb-doped TiO$_2$-terminated SrTiO$_3$ (001) substrate before
depositing the ferroelectric BaTiO$_3$ thin
film~\cite{Molegraaf-AdM-2009}.
 
\begin{figure}[t]
\includegraphics[angle=-90,width=\textwidth]{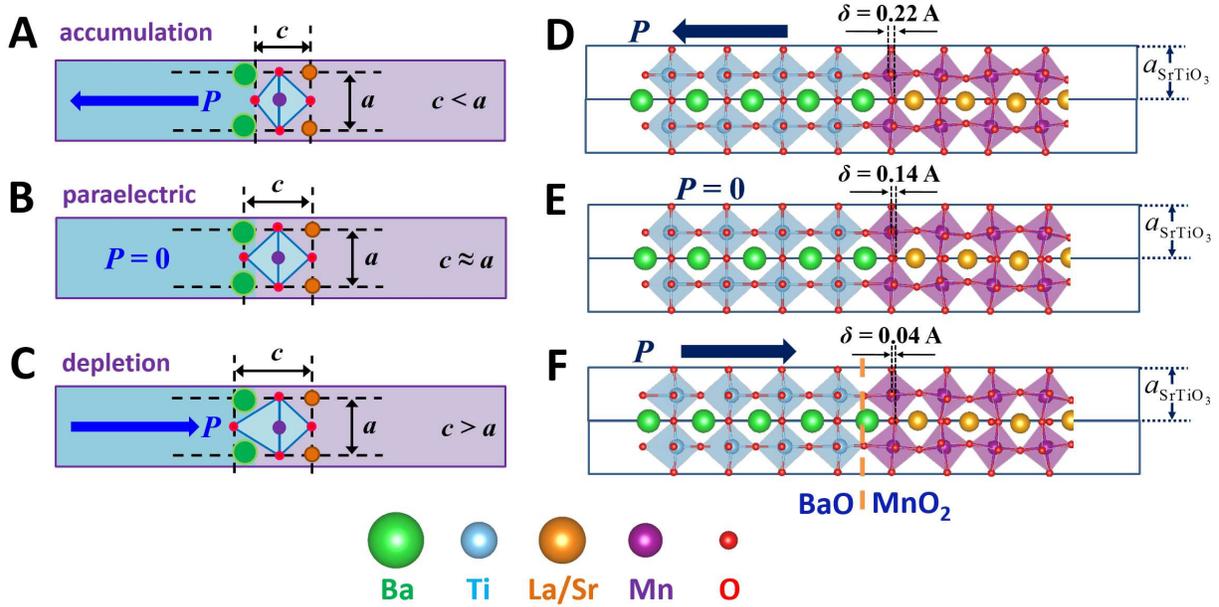}
\caption{\label{fig:supercell} Panels \textbf{A}-\textbf{C}:
Schematic of a BaTiO$_3$/LSMO interface. The purple part represents
LSMO and the light blue part is BaTiO$_3$. The interface is BaO/MnO$_2$.
The oxygen octahedron enclosing the interfacial Mn atom changes its
 $c/a$ ratio as the ferroelectric polarization flips.
\textbf{A}) Accumulation state; \textbf{B}) paraelectric state;
\textbf{C}) depletion state. Panels \textbf{D}-\textbf{F}: Relaxed
atomic structure of LSMO/BaTiO$_3$ interfaces from first-principles
calculations. The orange dashed line in \textbf{F} highlights the BaO/MnO$_2$
interface. The whole structure is strained to a SrTiO$_3$
substrate (substrate not shown in the figure). $\delta$ is the
Mn-O displacement. \textbf{D}) Accumulation state; \textbf{E})
paraelectric state; \textbf{F}) depletion state. }
\end{figure}

First principles calculations have shown that at a ferroelectric/LSMO
interface, the termination of the ferroelectric polarization and
presence of surface bound charge pull screening charges to the
interface~\cite{Duan-PRL-2006,Rondinelli-NatNanotech-2008,Chen-PRB-2012}.
For out-of-plane ferroelectric polarizations, two interfacial states are
possible: accumulation or depletion of holes, as illustrated in the
left panels of ~\ref{fig:supercell}. For accumulation, the
interfacial BaO layer is polarized with its O anion pushed towards the
interfacial Mn; due to the epitaxial constraint on the in-plane
lattice constant $a$, this means that the out-of-plane lattice
constant $c$ becomes smaller than $a$ so $c/a<1$ for the octahedral
oxygen cage surrounding the interfacial Mn, lending to stabilization
of the in-plane $d_{x^2-y^2}$ (as per standard crystal field
theory~\cite{VanVleck-PRB-1932}). For depletion, the BaO layer's
oxygen is pushed away from the Mn, leading to $c>a$ and favoring
$d_{3z^2-r^2}$. In addition, we compute an artificial ``paraelectric''
reference state where the BaTiO$_3$ is fixed to be non-polarized and
where we expect $c\approx a$.  The fully relaxed atomic structures
from first principles calculations are shown in the right panels of
~\ref{fig:supercell}.

\begin{figure}[t]
\includegraphics[angle=-90,width=0.8\textwidth]{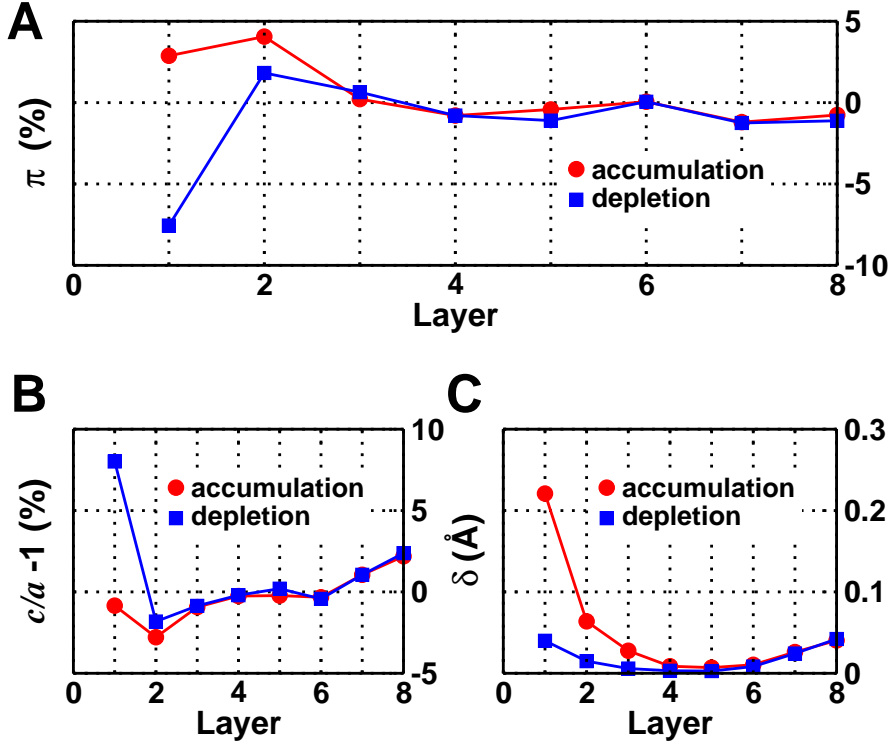}
\caption{\label{fig:orbital} \textbf{A}) Layer-resolved orbital
  polarization $\pi^i$ on successive Mn cations.  Layer 1 is at the interface.
\textbf{B}) $c/a$ ratio of each oxygen octahedron that encloses Mn atoms.
\textbf{C}) Mn-O displacement $\delta$ of successive MnO$_2$ layers.}
\end{figure}

To quantify the difference in orbital populations, we use a standard
definition of orbital polarization $\pi$ from
Ref.~\cite{Benckiser-NatMater-2011}
\begin{equation}
\label{equ:pol} \pi^i=\frac{n^i_{d_{x^2-y^2}} - n^i_{d_{3z^2-r^2}}}
{n^i_{d_{x^2-y^2}} + n^i_{d_{3z^2-r^2}}} =
\frac{\left(n^i_{d_{x^2-y^2}}/n^i_{d_{3z^2-r^2}}\right) - 1}{\left(n^i_{d_{x^2-y^2}}/n^i_{d_{3z^2-r^2}}\right) + 1}
\end{equation}
where $n^i_{\alpha}$ is the occupancy of atomic orbital $\alpha$ in
the $i$th unit cell of the LSMO. \ref{fig:orbital}\textbf{A} shows the
computed $\pi^i$ of each layer moving away from the interface. The
positive interfacial orbital polarization for accumulation means that
$d_{x^2-y^2}$ is stabilized while in depletion the $\pi^i<0$ means
that $d_{3z^2-r^2}$ is more populated. Qualitatively, these results
are consistent with our preceding schematics-based
expectations. However, the interfacial $\pi^i$ values shown in
~\ref{fig:orbital}\textbf{A} are hard to rationalize using the actual
$c/a$ ratios in~\ref{fig:orbital}\textbf{B} because for accumulation
$c/a$ is very close to unity, but we find significant positive
$\pi^i$.  The key neglected degree of freedom turns out to be the
polar distortion (ferroelectric displacement) of the interfacial
MnO$_2$ layer.  \ref{fig:orbital}\textbf{C} shows the displacement
amplitude $\delta$ in each MnO$_2$ layer~\cite{delta}.  While $\delta$
is small (0.04~\AA) for depletion, it is as large as 0.22~\AA~for
accumulation. There are at least two mechanisms that create the large
displacement for accumulation. First, $\delta$ is the continuation of
the ferroelectric distortion from the BaTiO$_3$ into the interfacial
layers, much like in PbTiO$_3$/SrTiO$_3$
superlattices~\cite{Dawber-PRL-2005}. Second, the paraelectric
reference calculation (\ref{fig:supercell}\textbf{E}) shows that
the interfacial MnO$_2$ layer, due to an asymmetric chemical
environment with BaO layer on one side and (La$_{1-x}$Sr$_x$)O layer
on the other side, has a significant $\delta$ even in the absence of
ferroelectricity, as also observed
elsewhere~\cite{Benckiser-NatMater-2011}. In accumulation, both
effects add and lead to a large displacement
(\ref{fig:supercell}\textbf{D}). In depletion, they oppose and
thus we have a much smaller displacement
(\ref{fig:supercell}\textbf{F}).

\begin{figure}[t]
\includegraphics[angle=-90,width=0.98\textwidth]{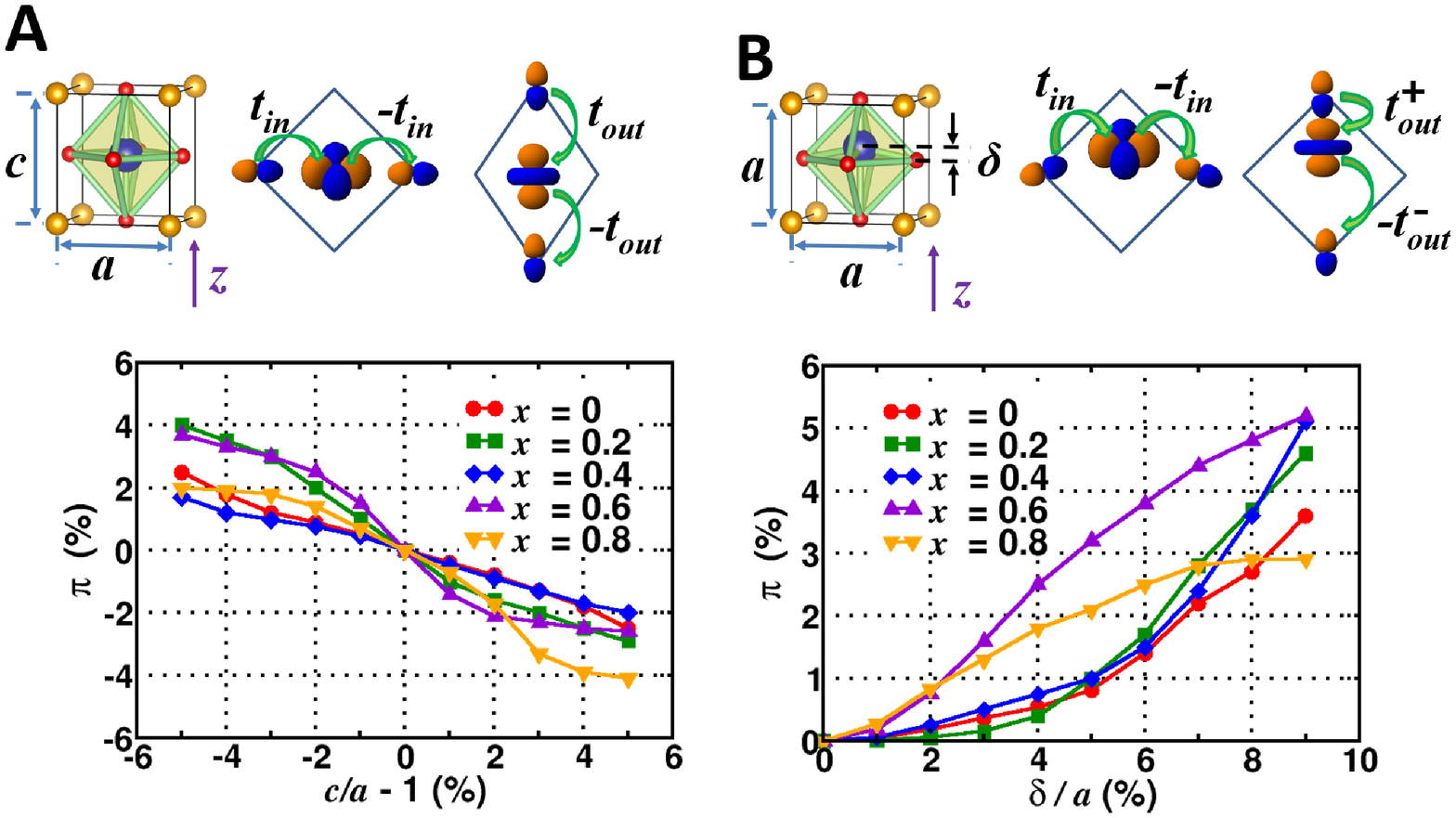}
\caption{\label{fig:bulk} Orbital polarization $\pi$ of bulk LSMO
as a function of \textbf{A}) $c/a$ ratio and \textbf{B}) Mn-O
displacement $\delta$ along the $z$ direction. $x$ is the hole doping of LSMO.
The in-plane (out-of-plane) Mn-O hopping is shown in schematics,
denoted by $t_{\textrm{in}}$ ($t_{\textrm{out}}$).}
\end{figure}

Does the Mn-O displacement $\delta$ create Mn orbital polarization as
effectively as the canonical Jahn-Teller-like $c/a$ distortions?
First principles computations allow us to separate these two effects
and study them separately. \ref{fig:bulk}\textbf{A} and
\textbf{B} show how the orbital polarization $\pi$ of bulk LSMO is
modulated independently by $c/a-1$ and $\delta$, respectively. Each
creates significant orbital polarization by itself.  For the
interfacial system, \ref{fig:orbital}\textbf{B} and
~\ref{fig:orbital}\textbf{C}
show that, in accumulation, $c/a$ is close to unity but $\delta$ is
large, so it must be $\delta$ that creates $\pi>0$ in accumulation. On
the other hand, in depletion, $c/a$ is significantly larger than unity
but $\delta$ is quite small, therefore it is $c/a>1$ that generates
$\pi<0$.

The underlying microscopic mechanisms for the $c/a$ and $\delta$
dependence of $\pi$ follow from considering the modifications of
nearest neighbor hoppings between Mn and O at the interface as shown
schematically in~\ref{fig:bulk}\textbf{A} and \textbf{B}. For the $c/a$
dependence, we begin with the fact that, in the metal oxide, the two
Mn $e_g$ states are anti-bonding in nature. A larger (smaller) Mn-O
hopping, denoted by $t$, leads to a higher (lower) center-of-band
energy. Therefore, if $c/a > 1$ ($c/a < 1$), the in-plane
(out-of-plane) hopping $t$ is larger and thus the $d_{x^2-y^2}$
($d_{3z^2-r^2}$) orbital has higher energy and less
occupancy. Therefore, $c/a > 1$ ($c/a < 1$) leads to $\pi < 0$ ($\pi >
0$).

The $\delta$ dependence is different in nature as $\delta\ne0$ always
stabilizes $d_{x^2-y^2}$ and leads to $\pi>0$. The reason is that
$\delta\ne0$ breaks the inversion symmetry of the MnO$_6$ octahedron
which means that there are two different out-of-plane hoppings
$t^{+}_{\textrm{out}} > t^{-}_{\textrm{out}}$ (see~\ref{fig:bulk}\textbf{B}). The broken symmetry and differing hoppings
permit mixing of the $d_{3z^2-r^2}$ and apical O $p_z$ orbitals at the
conduction band edge at the $\Gamma$ point, a forbidden mixing when
inversion symmetry is present. This makes for a more anti-bonding
$d_{3z^2-r^2}$ orbital that is pushed to higher energy and
consequently is less occupied. A more detailed explanation can be found in
the Supplemental Material~\cite{supplement}.

\begin{figure}[t]
\includegraphics[angle=-90,width=10cm]{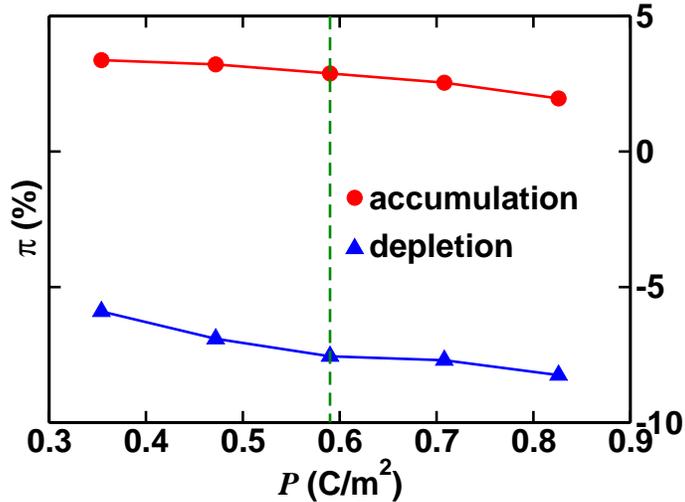}
\caption{\label{fig:pol_dep} Dependence of interfacial Mn orbital
polarization $\pi$ as a function of ferroelectric polarization
$P$. The green dashed lines correspond to the calculated
ferroelectric polarization of SrTiO$_3$-strained bulk BaTiO$_3$.}
\end{figure}

The above results have used BaTiO$_3$ as a prototype ferroelectric, but
the reversible nature of the modulated interfacial orbital
polarization is generic in that it does not depend on a specific
choice of ferroelectric.  First, we verify the robustness of
our prediction by manually changing the ferroelectric polarization
magnitude over a wide range from 0.35 to 0.83 C/m$^2$
(SrTiO$_3$-strained bulk BaTiO$_3$ has a theoretical polarization of
0.59 C/m$^2$, consistent with previous studies~\cite{Bilc-PRB-2008, 
Neaton-MRS-2002, Ederer-PRL-2005}).
\ref{fig:pol_dep} shows that the orbital polarization remains
significant with a difference of around 10\% between the two
polarization states. This difference should be detectable by x-ray
linear dichroism (XLD)~\cite{Tebano-PRL-2008, Huijben-PRB-2008} or
orbital reflectometry~\cite{Benckiser-NatMater-2011}. Second, we
have explicitly computed and compared tetragonal BaTiO$_3$/LSMO and 
PbTiO$_3$/LSMO interfaces, (see Figure~S2 and Figure~S3 in the
Supplemental Material). The resulting orbital polarization of the
PbTiO$_3$/LSMO interface is qualitatively similar to that of the
BaTiO$_3$/LSMO interface, but is quantitatively larger than 
the orbital polarization of the BaTiO$_3$/LSMO interface 
due to the larger ferroelectric polarization of PbTiO$_3$.

A key prediction from our calculations is the significant Mn-O
displacement $\delta$ in \textit{metallic} manganites in the
accumulation state. This polar distortion is a genuine interfacial
phenomenon and stems from the propagation of ferroelectric
polarization and a \textit{finite} screening length of
LSMO~\cite{Salamon-RMP-2001}. To verify the theoretical predictions,
we combine thin film growth techniques and electron microscopy to
characterize the atomic-scale geometry and electronic structure of
ferroelectric/manganite interfaces. We show, experimentally, that the
screening length of LSMO is about 12~\AA~(i.e. 3 unit cells) and a
Mn-O displacement does exist at the interface, the magnitude of which
is in good agreement with our theoretical prediction.

\begin{figure}[h!]
\includegraphics[angle=-90,width=0.85\textwidth]{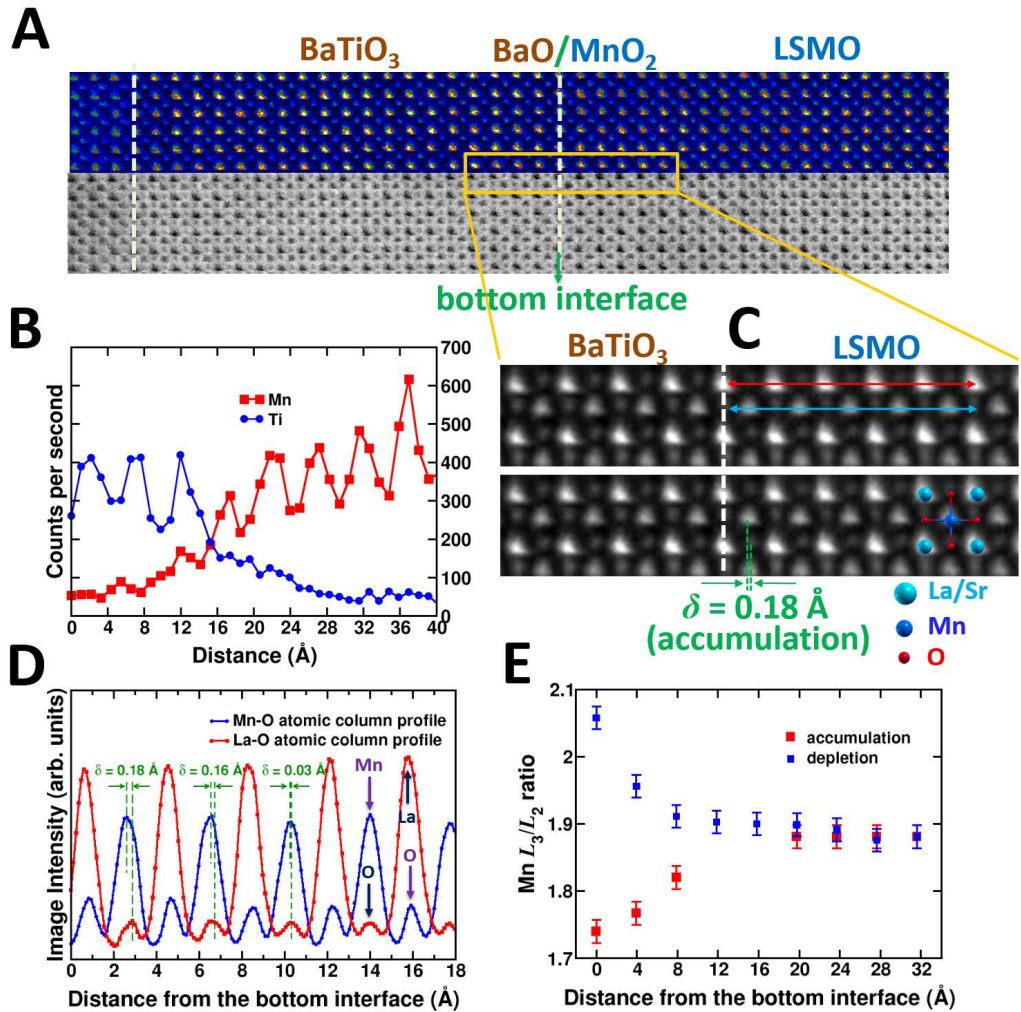}
\caption{\label{fig:EELS} \textbf{A}) HAADF/ABF images of the
  LSMO/BaTiO$_3$/LSMO film with the interfaces marked and O columns
  clearly visible in ABF; \textbf{B}) an EDX line scan acquired along
  the transition metal column across the interface, demonstrating its
  atomically-sharp nature. The red and blue symbols are for Mn $K$-series and 
  Ti $K$-series x-rays, respectively. \textbf{C}) interpolated, filtered,
  inverted, and averaged ABF image of the bottom interface
  (highlighted by the yellow box) which reveals the Mn-O displacements
  in the accumulation state. The apparent triangular shape of the
  atomic columns is an artifact of the cross-correlation and averaging
  process. The red and blue solid lines highlight the La-O and Mn-O
  column profiles; \textbf{D}) intensity line profiles of the Mn-O and
  La-O atomic column revealing the Mn-O displacements as shown
  theoretically in~\ref{fig:supercell}\textbf{D}, highlighted
  with green lines are the displaced Mn and O planes.  The higher
  peaks in the two curves correspond to the positions of La and Mn
  atoms and the lower peaks to the positions of O atoms; \textbf{E})
  the measured Mn $L_3$/$L_2$-ratio for the accumulation (depletion)
  state as a function of distance into LSMO from the bottom interface
  (the measurements are done in two separate samples).}
\end{figure}

Observation of atomic-scale structural distortions at the interface
between manganites and ferroelectrics necessitates atomically abrupt
interfaces. To experimentally realize atomically abrupt interfaces, we
use molecular beam epitaxy (MBE) to grow LSMO and BaTiO$_3$, which
have similar growth conditions and can be grown in the same MBE
chamber. As such, we grow LSMO/BaTiO$_3$/LSMO heterostructures on
SrTiO$_3$(001) substrates, as described in the Supplemental
Material~\cite{supplement}. 
Throughout this study all transmission electron microscopy
(TEM) measurements are performed at the bottom interface only
(i.e. the interface closer to the substrate, which is highlighted by
the yellow box in~\ref{fig:EELS}\textbf{A}) and has a BaO/MnO$_2$
termination (consistent with~\ref{fig:supercell}).  By analyzing
only the bottom interface, we ensure that there are no variations in
stoichiometry that may occur during sample growth. Two different TEM
samples are prepared from the same as-grown wafer, and each TEM sample
is found to exhibit different directions of the ferroelectric
polarization (accumulation and depletion, respectively). These
different polarization directions may have arisen during TEM sample
preparation, but facilitate analysis of the LSMO. From now on, the
sample with the bottom interface in the accumulation state is called
`accumulation sample' and the other with the bottom interface in the
depletion state is called `depletion sample'. We characterize both TEM
samples via aberration-corrected scanning transmission electron
microscopy (STEM) to identify the atomic and electronic structure of
the bottom interface. All high-angle annular dark-field
(HAADF)~\cite{Pennycook-Nature-1988} and annular bright field
(ABF)~\cite{Findlay-APL-2009} STEM images and electron energy loss
(EEL) spectra are acquired on a probe-corrected JEOL
JEM-ARM200CF~\cite{JEOL} operated at 200 kV with a 22 mrad convergence
angle.

Since the theoretically predicted Mn-O displacement in
\textit{metallic} manganites is found in the accumulation state, we
first focus on the accumulation sample. \ref{fig:EELS}\textbf{A}
shows a pair of HAADF and ABF images of the entire accumulation sample
(LSMO/BaTiO$_3$/LSMO thin film in the [001] direction), along with
pertinent EEL spectra.  The bottom interface, which is highlighted by
the yellow box in~\ref{fig:EELS}\textbf{A}, is in the
accumulation state. The HAADF image shows a largely defect-free,
atomically abrupt BaTiO$_3$/LSMO interface. Within the BaTiO$_3$, the
Ba atoms appear as the brighter spots in HAADF forming a rectangular
lattice with the Ti located at the center of these rectangles. O
atomic columns are observed in the ABF image~\cite{Kim-ACSnano-2013}.
The position of the interface and the interfacial sharpness were
investigated using energy-dispersive x-ray spectroscopy (EDX), with an
example line scan shown in~\ref{fig:EELS}\textbf{B}.  When the
electron probe is scanned along the transition metal column, it is
clear that the interface is atomically-sharp within one unit cell.
The remaining intensity both on and off the scanned column is likely
an artifact of the probe dechanneling and not due to diffusion or
interfacial roughness. The details concerning the EDX parameters can
be found in the Supplemental
Material~\cite{supplement}. \ref{fig:EELS}\textbf{C} is an
inverted and averaged ABF image -- previously interpolated ($2\times$)
and filtered -- of the bottom interface (highlighted by the yellow box
in~\ref{fig:EELS}\textbf{A}), which reveals the Mn-O displacement
$\delta$ in the first few MnO$_2$ layers. The magnitude of the Ti-O
displacement in the BaTiO$_3$ is measured to be $\approx$ 0.13~\AA~
based on the inverted ABF image. \ref{fig:EELS}\textbf{D} shows
the Mn-O and La-O atomic column profiles (more precisely, the La here
refers to La$_{1-x}$Sr$_x$), which are fit with a Gaussian. The peak
positions of Mn and La (from the Mn-O and La-O line profiles) are
determined from the Gaussian fit: we find $\delta$ of 0.18~\AA,
0.16~\AA~and 0.03~\AA~in the first three layers of MnO$_2$ from the
bottom interface (highlighted by the yellow box in
~\ref{fig:EELS}\textbf{A}). In subsequent MnO$_2$ layers, $\delta$
is below the measurement limit. The theoretically predicted Mn-O
displacements at the bottom interface in the accumulation state are
clearly evidenced in the atomic column profile images with correct
sign albeit reduced value. The reduction may result from imaging
artifacts due to the convolution of the Mn and O peaks in the mixed
Mn-O column and the fact that GGA-PBE overestimates the theoretical
ferroelectric polarization of BaTiO$_3$~\cite{Bilc-PRB-2008}. The Mn-O
displacements observed in experiment are further quantitatively
confirmed by calculating the ion displacements from the multi-slice
simulated STEM images of the optimized theoretical structures which
also yields interfacial Mn-O displacements of a smaller magnitude than
\textit{ab initio} values (see Figure~S9 in the Supplemental Material
for details~\cite{supplement}).

Next, we measure the bottom interface in both accumulation and
depletion samples. One- and two-dimensional atomically-resolved EEL
spectra data are acquired from the bottom interface and the bulk-like
regions of the LSMO~\cite{EELS}. \ref{fig:EELS}\textbf{E} shows
the Mn $L_3/L_2$ ratio as a function of distance into the LSMO from
the bottom interface in the accumulation state and in the depletion
state, respectively (please see Section VI in the Supplemental
Material~\cite{supplement} for details on Mn $L$-edge EELS).  The
change of Mn $L_3/L_2$ ratio in the conducting LSMO directly reveals the
collection of mobile screening charges that electrostatically screen
the ferroelectric polarization at the interface. We observe an excess
of screening holes for accumulation (electrons for depletion) that
decays back to the bulk LSMO level over about three unit cells, which
is a measure of the \textit{finite} screening length of metallic
LSMO.

To summarize the experiments, STEM images and EEL spectra
separately corroborate the intended polarization and the presence
of an accumulation or depletion region at the interface. A finite
screening length about 12~\AA~for metallic LSMO is verified via the Mn
$L_3/L_2$ ratio. More importantly, the theoretically predicted interfacial 
Mn-O displacement of the accumulation state in \textit{metallic} 
manganites is directly observed in our experimental HAADF/ABF images.

Before we conclude, we comment that though the crucial Mn-O
displacement at the interface is observed in experiment, a direct
measurement of orbital polarization is more desirable to confirm our
theoretical predictions. Such an x-ray dichroism measurement at a
buried interface of a complicated heterostructure is experimentally
challenging, and is an active area of research. However, the
switchable polarization of ferroelectrics via external field yields a
reversible orbital polarization at the ferroelectric/manganites
interface, which is easier to detect, since by measuring the
change in the signal of x-ray dichroism, any potential difficulties
due to a bulk-like background are automatically eliminated.
Therefore, we hope that our theoretical results stimulate further
experiments that explore revserible control of orbital degree of
freedom, in addition to the control of charge and spin in transition metal
oxides~\cite{Vaz-PRL-2010}, and functional catalysis at ferroelectric
interfaces and surfaces~\cite{Li-NatMat-2008, Kim-PRL-2011}.

In conclusion, we have used \textit{ab initio} calculations
combined with experimental growth and characterization to describe
the atomic-scale geometry and electronic structure of
ferroelectric/manganite interfaces. The orbital degeneracy of the
Mn $e_g$ states at such an interface is removed in a reversible
manner: by changing the ferroelectric polarization, one can change
the sign and magnitude of the orbital degeneracy breaking.
Microscopically, a new structural distortion, absent in bulk
manganites, is shown to be critical in determining the orbital
polarization: the polar displacement of the interfacial
cation-anion layer. Moreover, this mechanism is generic and should
be present in other interfacial materials systems.

\section{Supporting Information}
Details on \textit{ab initio} calculations, tight-binding analysis, 
thin film growth and characterization. This material is available 
free of charge via the Internet at http://pubs.acs.org.

\begin{acknowledgement}
This work at Yale was supported by NSF MRSEC DMR 1119826, NSF CNS
08-21132, FAME, ONR, and by the facilities and staff of the Yale
University Faculty of Arts and Sciences High Performance Computing
Center.  Additional computations used the NSF XSEDE resources via
grant No. TG-MCA08X007 and No. TG-PHY130003. At UIC, the work was
supported by a grant from the National Science Foundation
[DMR-0846748]. The acquisition of the UIC JEOL JEM-ARM200CF is
supported by a MRI-R$^2$ grant from the National Science Foundation
[DMR-0959470]. Support from the UIC Research Resources Center is also
acknowledged.
\end{acknowledgement}

\providecommand*{\mcitethebibliography}{\thebibliography}
\csname @ifundefined\endcsname{endmcitethebibliography}
{\let\endmcitethebibliography\endthebibliography}{}


\begin{mcitethebibliography}{41}
\bibitem[del()]{delta}
 Since the MnO$_6$ of La$_{1-x}$Sr$_x$MnO$_3$ has rotation and tilting, we
  calculate the Mn-O displacement along the [001] direction (i.e $z$-axis).
  Then we average the Mn-O displacements of the two MnO$_6$ octahedra in each
  manganite layer.\relax
\mciteBstWouldAddEndPunctfalse
\mciteSetBstMidEndSepPunct{\mcitedefaultmidpunct}
{}{\mcitedefaultseppunct}\relax
\EndOfBibitem
\end{mcitethebibliography}

\end{document}